\documentclass[prl,twocolumn, amsmath,amssymb,
reprint,%
author-year,%
author-numerical,%
dvipdfmx
]{revtex4-1}

\usepackage{graphicx}
\usepackage{bm}
\usepackage{color}

\begin{document}


\title{Stability of B2 compounds: Role of the $M$ point phonons}

\author{Shota Ono}
\email{shota\_o@gifu-u.ac.jp}
\author{Daigo Kobayashi}
\affiliation{Department of Electrical, Electronic and Computer Engineering, Gifu University, Gifu 501-1193, Japan}

\begin{abstract}
Although many binary compounds have the B2 (CsCl-type) structure in the thermodynamic phase diagram, an origin of the structural stability is not understood well. Here, we focus on 416 compounds in the B2 structure extracted from the Materials Project, and study the dynamical stability of those compounds from first principles. We demonstrate that the B2 phase stability lies in whether the lowest frequency phonon at the $M$ point in the Brillouin zone is endowed with a positive frequency. We show that the interatomic interactions up to the fourth nearest neighbor atoms are necessary for stabilizing such phonon modes, which should determine the minimum cutoff radius for constructing the interatomic potentials of binary compounds with guaranteed accuracy.
\end{abstract}

\maketitle


{\it Introduction.} Since the characterization of the crystal structure of CsCl in 1921 \cite{davey}, many of the B2 (called CsCl-type) compounds have been synthesized experimentally, resulting in that the B2 structure is the most common phase in the thermodynamic phase diagram of binary compounds. Analyses of materials database have shown how frequent they appear \cite{sluiter,hart2007,kolli}. More recently, Kolli {\it et al}. identified 267 parent crystal structures that can generate their derivative ordered phases, and showed that the body-centered cubic (bcc) structure is the most common parent crystal structure \cite{kolli} by using the Materials Project (MP) database \cite{MP}. Among them, the B2 structure is the most common ordering on bcc structure. Apart from the ambient condition, the B2 structure may also appear: the B2 structure can be transformed from the B1 (NaCl-type) structure under high pressure (such as alkali halides and alkali-earth oxides \cite{sims,florez}) and from the L1$_0$ (CuAu-type) structure in warm dense matter regime \cite{ono_koba}. 


Although the elastic stability of the B2 structure has been studied in a wide variety of compounds \cite{khenata,wangB2,alsalmi}, such a stability does not always yield the dynamical stability against the zone boundary phonon excitations \cite{grimvall}. For the parent bcc structure, it is well known that the transverse acoustic phonon at the $N$ point in the Brillouin zone (BZ), propagating along the diagonal direction of any two axes in the cubic cell, has relatively low frequencies \cite{frank,persson}. It has been shown that such a phonon is stabilized by long-range interatomic interactions, allowing alkali metals to form the bcc structure at the ambient condition \cite{ono2019}. By considering the fact that the B2 structure is equivalent to the bcc structure when two species are assumed to be the same element, we expect that a similar scenario holds: the lowest frequency phonon at the $M$ point that is stabilized by the long-range interactions determines the stability of the B2 compounds. We have recently confirmed that the long-range interatomic interactions up to the fifth nearest neighbor (5NN) atoms are needed to understand the dynamical stability of the CuAu in the L1$_0$ structure \cite{ono_koba}. 

The range of the interatomic interactions is of prime importance in the field of atomistic modeling of condensed matters. For the bcc elemental metals, the cutoff radii should be larger than the 3NN or 4NN distances \cite{adam,seko,zuo}. However, it has not been understood why more than 3NN distances are required to describe the potential energy surface accurately. 

In this paper, we study the dynamical stability of 416 compounds in the B2 structure from first principles, by assuming zero temperature and pressure. We show that 266 out of 416 compounds are dynamically stable, and demonstrate that the stability of the B2 compounds is mainly determined by whether the lowest frequency phonon at the $M$ point in the BZ is endowed with a positive frequency. In addition, we develop a force constant model taking into account the interatomic interactions up to the 6NN atoms, and demonstrate that the interatomic interactions up to the 4NN atoms are enough to stabilize the lowest energy phonons at the $M$ point. This should determine a minimum cutoff radius of the interatomic potentials for binary compounds. The present work unveils the microscopic mechanism of the B2 compounds stability, which stimulates other studies involving different crystal structures. 


\begin{figure*}
\center
\includegraphics[scale=0.53]{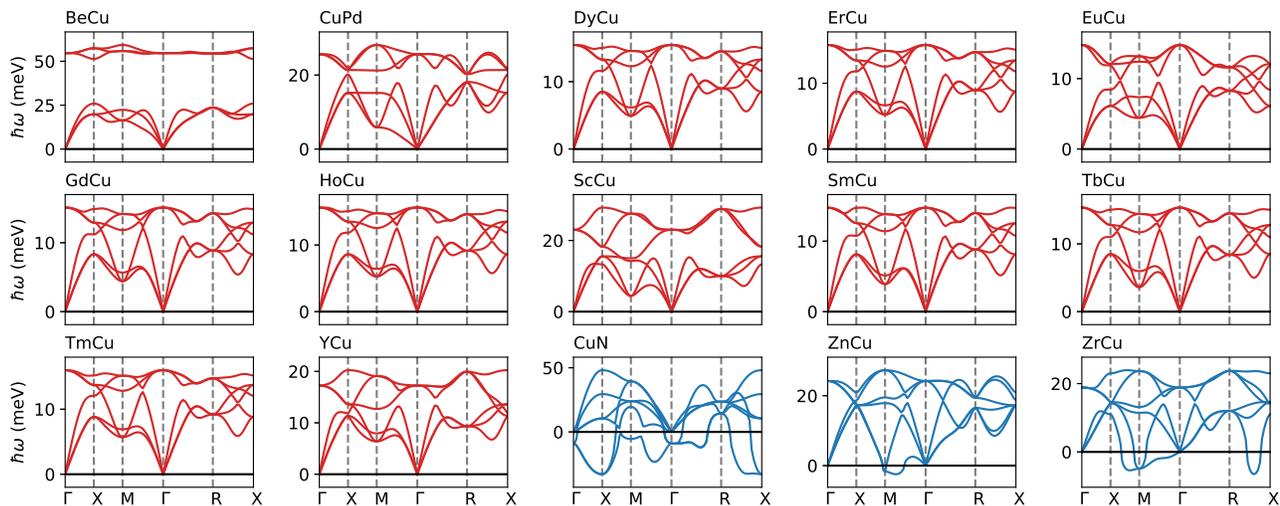}
\caption{The phonon dispersions of the Cu-based B2 compounds. The dispersion curves are colored red and blue for the dynamically stable and unstable compounds, respectively. } \label{fig1} 
\end{figure*}

{\it Methods.} By using the MP \cite{MP} and the \texttt{pymatgen} \cite{pymatgen}, we first extracted the list of the B2 compounds. By setting the space group to $Pm\bar{3}m$ with the number of atoms in a unit cell being two and excluding the atoms in the actinide series, 416 B2 compounds having the inorganic crystal structure database (ICSD) IDs were found. 

We next optimized the lattice parameter $a$ for the 416 compounds. All the density-functional theory (DFT) calculations were performed with the \texttt{Quantum ESPRESSO} (QE) \cite{qe} using the Perdew-Burke-Ernzerhof (PBE) \cite{pbe} functionals of the generalized gradient approximation for the exchange-correlation energy and using the ultrasoft pseudopotentials provided in the \texttt{pslibrary.1.0.0} \cite{dalcorso}. Spin-polarized approximation was used for all calculations. The cutoff energies for the wavefunction and the charge density are 60 Ry and 600 Ry, respectively, and 20$\times$20$\times$20 $k$ grid was used in the self-consistent field (SCF) calculations \cite{MK}. The SCF convergence threshold was set to be $10^{-8}$ Ry and the smearing parameter of $\sigma=0.02$ Ry \cite{smearing} was used for all calculations. The total energy and forces were converged within $10^{-5}$ Ry and $10^{-4}$ a.u., respectively. 

The accuracy of the present calculations was checked by comparing the optimized $a$'s with those in the MP \cite{MP}. We have confirmed that the optimized $a$'s agree with the reference values within an error of 1 \% except for the 11 Ce-based compounds, ClO, and NCl. The optimized $a$'s in the Ce-based compounds are larger than the reference values by a few percent. This may be due to the absence of the $f$-electrons in the present calculations, resulting in no magnetic moments, whereas the Ce-based compounds show ferromagnetic phase in the MP \cite{MP}. The error of the $a$'s for ClO and NCl were 4.6 and 11.1 \%, respectively. 

Although the formation energies $E_{\rm form}$ of the 416 compounds can be obtained from the MP \cite{MP}, the dynamical stability properties are not always obtained. We thus performed phonon dispersion calculations based on density-functional perturbation theory (DFPT) \cite{dfpt} implemented in QE. The threshold parameter for the self-consistency (tr2\_ph) was set to be $10^{-14}$, and 4$\times$4$\times$4 $q$ grid (10 $q$ points) was used. We calculated the phonon dispersions along the symmetry lines $\Gamma$-$X$-$M$-$\Gamma$-$R$-$X$. When the phonon frequency $\omega$ is imaginary, the phonon energy is represented as a negative value, $-\hbar\vert\omega\vert$, with the Planck constant $\hbar$. In the present work that adapts a finite size $q$ grid, we identify the B2 compound as dynamically stable if the lowest phonon energy is larger than $\varepsilon_{\rm min}=-1$ meV. 

\begin{figure}
\center
\includegraphics[scale=0.42]{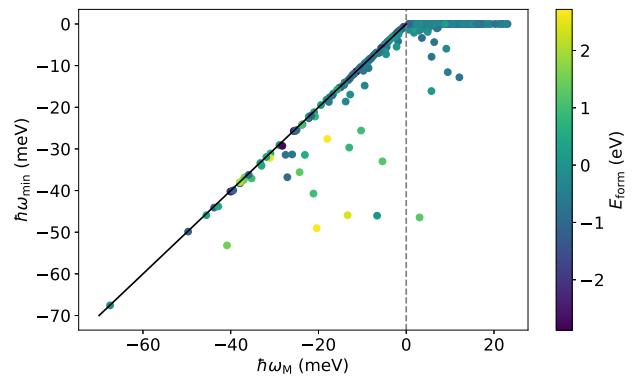}
\caption{The minimum phonon energy plotted as the $M$ point phonon energy. The solid line indicates the relation of $\omega_{\rm M}=\omega_{\rm min}$. The color of the data points indicates the value of the formation energy extracted from the MP database \cite{MP}. } \label{fig2} 
\end{figure}

{\it Phonon dispersions.} We have found that 266 out of 416 compounds are identified to be dynamically stable. For example, we show the phonon dispersions for the 15 Cu-based compounds in Fig~\ref{fig1}. The 12 compounds (see Fig.~\ref{fig1}) out of 15 are dynamically stable, whereas the others (CuN, ZnCu, and ZrCu) are unstable. Although the former 12 compounds are stable, the phonon energy of the transverse acoustic branch at the $M$ point tends to be small, compared to that at the other symmetry points (except for $\Gamma$) in the BZ. In a similar manner, the instabilities of the ZnCu and ZrCu are due to the phonon softening around the $M$ point, leading to negative phonon energies. The CuN shows strong phonon softening over the entire BZ. This may be ascribed to the positive formation energy (1.22 eV/atom \cite{MP}), implying that the B2 CuN is decomposed into the Cu crystal in the face-centered cubic structure and the N$_2$ molecules. 

For the B2 compounds including the platinum-group metals (Os, Ru, Ir, Rh, Pt, and Pd), Hart {\it et al}. have identified that 16 compounds have the B2 structure as the ground state by using the DFT calculations within the PBE \cite{hart}. We have also identified that the 16 compounds are dynamically stable. In addition, we find that the B2 FeRh is dynamically stable, which is consistent with the experimental synthesis \cite{FeRh}, whereas the FeRh in the B2 structure has been predicted to be unstable \cite{hart}. 

Overall, the dynamical stability of the B2 compounds is determined by the $M$ point phonon energy. The phonon dispersions for the 416 compounds are shown in the Supplemental Material \cite{SM}. 

{\it The $M$ point phonons.} In order to understand to what extent the $M$ point phonon determines the dynamical stability of the B2 compounds, we plot the minimum phonon energy within the entire BZ ($\hbar\omega_{\rm min}$) as a function of the lowest phonon energy at the $M$ point ($\hbar\omega_{\rm M}$) in Fig.~\ref{fig2}. When $\omega_{\rm M}\le 0$, the relationship of $\omega_{\rm M}=\omega_{\rm min}$ holds, whereas when $\omega_{\rm M}>0$, the equality of $\omega_{\rm min}=0$ holds. These show that the stability of the $M$ point phonon is a good descriptor for identifying the dynamical stability of the B2 compounds. 
 
In Fig.~\ref{fig2}, we can find some exceptions that do not follow the B2 stability criteria above. There are 16 compounds (AlPt, CsN, KS, LuMg, MgHg, MgTl, MnAu, MnHg, NaS, PrMg, SiRh, TcB, TiRh, VRu, YMg, and ZrRh) that satisfy both $\omega_{\rm M}>0$ and $\hbar\omega_{\rm min}<\varepsilon_{\rm min}$. Although the $M$ point phonon does not have the minimum phonon energy in these compounds, negative phonon energies appear along the $\Gamma$-$M$ line in the BZ (see the Supplemental Material \cite{SM}). The instability of VRu is anomalous because negative phonon energies are observed only along the $\Gamma$-$R$ line in the BZ, while the experimental synthesis of the VRu in the B2 and tetragonally distorted B2 structures \cite{VRu1,VRu2} has been reported. 

The value of $E_{\rm form}$ may be another descriptor for understanding the stability of the B2 compounds: the dynamically stable compounds have negative $E_{\rm form}$ and, in turn, the unstable compounds have positive $E_{\rm form}$. However, many exceptions have been found: (i) the six compounds of CaNi (0.024), CeMg (0.087), CrCo (0.154), LiBe (0.366), MnZn (0.062), and YbRu (0.209) have positive $E_{\rm form}$ (eV/atom), where the figure in a parenthesis indicates the magnitude of $E_{\rm form}$, although they are dynamically stable. In contrast, (ii) 103 out of 150 unstable compounds have negative $E_{\rm form}$: for example, CsF ($-2.734$), LiF ($-2.878$), RbF ($-2.775$), and SrO ($-2.662$) for strongly bonded systems and AlRe ($-0.010$), CrN ($-0.013$), and MnAu ($-0.010$) for weakly bonded systems. A similar issue has been found in a wide variety of materials, such as ordered alloys \cite{sun,ono_meta} and many two-dimensional materials \cite{wang,ono_simple,ono_satomi}. The metastability of materials has to be studied in detail. 

It is noteworthy that the B2 compounds including a semiconducting element (group 14-17) tend to be unstable (see Table S1 and Figure S1 \cite{SM}). Such compounds include the strongly bonded systems mentioned above. The instability of these may be due to the presence of different ground state structure such as the B1 structure. The understanding for the group dependence of the dynamical stability is left for future work. 

\begin{figure}
\center
\includegraphics[scale=0.5]{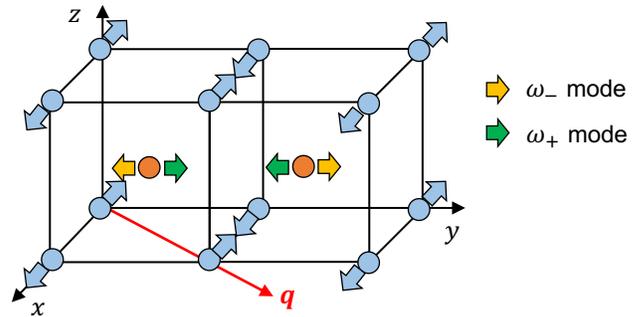}
\caption{The displacement vectors of the $M$ point phonon modes having the frequencies of $\omega_{\pm}$ (doubly degenerated). The cases that the atoms $A$ and $B$ move along the $x$ and $y$ directions, respectively, are illustrated. } \label{fig3} 
\end{figure}

\begin{figure*}
\center
\includegraphics[scale=0.48]{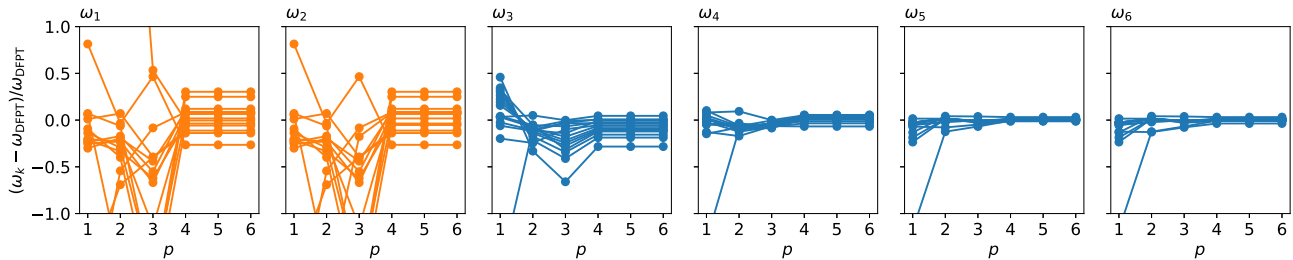}
\caption{The variation of the relative error of the $M$ point phonon energies between the $p$NN models and the DFPT for the 15 Cu-based compounds.} \label{fig4} 
\end{figure*}

{\it Effect of long-range interatomic interactions.} We next study an origin of the positive value for the phonon energies at the $M$ point and identify the role of the long-range interatomic interactions. Based on standard lattice dynamics \cite{maradudin}, we derived analytical expressions for the phonon frequencies for the wavevector $\bm{q}=(\pi/a,\pi/a,0)$, at the $M$ point in the BZ, by assuming the B2 compounds that consist of atoms $\kappa=A$ and $B$ with the masses $M_\kappa$. By using the translational symmetry of the crystal, the dynamical matrix $\tilde{D}$ is given by
\begin{eqnarray}
\tilde{D}_{\alpha\beta}^{\kappa\kappa'}(\bm{q})
 = \frac{1}{\sqrt{M_\kappa M_{\kappa'}}}
 \sum_{j} D_{\alpha\beta}^{\kappa\kappa'}(\bm{R}_j,\bm{0}) 
 e^{-i\bm{q}\cdot \bm{R}_j},
 \label{eq:dyn}
\end{eqnarray}
where $D_{\alpha\beta}^{\kappa\kappa'}(\bm{R}_j,\bm{0})$ is the force constant matrix, i.e., the force along the direction of $\alpha$ acting on the atom $\kappa$ in a unit cell characterized by the lattice vector $\bm{R}_j$ when the atom $\kappa'$ in the cell of $\bm{R}=\bm{0}$ is displaced along the direction of $\beta$. With the $q$ grid used in the present work, the $\bm{R}_j$ can take the vectors of $\sum_{i=1}^{3} m_i\bm{a}_i$ with $m_i = -1,0,1,2$, where $\bm{a}_i$'s are the primitive lattice vectors in the cubic cell. We thus considered the force constants up to the 6NN sites of B2 $AB$. By assuming that the atom $A$ is located at the origin, the position of the first, second, third, fourth, fifth, and sixth NNs are $B(a/2,a/2,a/2)$, $A(a,0,0)$, $A(a,a,0)$, $B(3a/2,a/2,a/2)$, $A(a,a,a)$, and $A(2a,0,0)$, respectively, and these equivalent sites, where the numbers of the equivalent sites are 8, 6, 12, 24, 8, and 6, respectively. Due to the equivalence between the $x$ and $y$ directions, four different phonon modes are present: the $z$-polarized modes having $\omega_{z\kappa}^{2}= \tilde{D}_{zz}^{\kappa\kappa}(\bm{q})$ ($\kappa=$ A or B) and the $x$- and $y$-polarized modes having 
\begin{eqnarray}
\omega_{\pm}^{2}= \frac{1}{2}\left[ \tilde{D}_{xx}^{AA}(\bm{q})+\tilde{D}_{yy}^{BB}(\bm{q}) 
\pm \tilde{C}_{xy}(\bm{q}) \right],
\label{eq:omegaI}
\end{eqnarray}
with 
\begin{eqnarray}
\tilde{C}_{\alpha\beta}(\bm{q}) &=& \sqrt{\left[\tilde{D}_{\alpha\alpha}^{AA}(\bm{q})
-\tilde{D}_{\beta\beta}^{BB}(\bm{q})\right]^2
+\left[2\tilde{D}_{\alpha\beta}^{AB}(\bm{q})\right]^2}.
\label{eq:Cq}
\end{eqnarray}
The value of $\omega_{\pm}^{2}$s in Eq.~(\ref{eq:omegaI}) does not change when the indexes $x$ and $y$ are replaced with $y$ and $x$, respectively, leading to the doubly degenerated modes. The displacement patterns of the normal modes with $\omega_{\pm}$ are described by a combination of the $x \ (y)$-polarized vibration of the atom $A$ and the $y \ (x)$-polarized vibration of the atom $B$, which are shown in Fig.~\ref{fig3}. From the expressions of Eqs.~(\ref{eq:omegaI}) and (\ref{eq:Cq}), one can expect that the coupling term of $\tilde{D}_{\alpha\beta}^{AB}(\bm{q})$ plays an important role to yield the positive value of $\omega_{-}^{2}$. The derivations of these expressions are provided in the Supplemental Material \cite{SM}. 

To study how the $M$ point phonons are stabilized by the interatomic interactions, we introduce the $p$NN model taking into account the force constants up to the $p$th NN atoms and compare the $M$ point phonon energies with the DFPT results. The errors of the phonon energies with restricted interactions, with respect to the DFPT results, are plotted as a function of $p$ in Fig.~\ref{fig4}. For clarity, the cases of the 15 Cu-based compounds are shown. The phonon frequencies are denoted as $\omega_k$ with $k=1,\cdots, 6$ in an ascending order. The lowest energy phonons ($\omega_1$ and $\omega_2$) correspond to the normal modes having the frequency $\omega_{-}$ except for CuN. For the low energy phonon modes, the deviation from the DFPT is large when $p\le 3$. The $p=4$ is a critical value for the convergence of the phonon energy, and such a $p$ corresponds to the interatomic interactions between different atoms $A$ and $B$, which is consistent with our expectation above. On the other hand, $p=2$ is found to suffice for the high energy phonons. These results indicate that the low and high energy phonons at the $M$ point are stabilized by the long-range and short-range interatomic interactions, respectively. The instability of the $M$ point phonon modes in the CuN was due to the negative value of $\omega_{z\kappa}^{2}= \tilde{D}_{zz}^{\kappa\kappa}(\bm{q})$ with $\kappa=$ Cu.  

The interatomic distance with the $p=4$ atoms is $\sqrt{11}a/2\simeq 1.66a$, and the total number of the NN atoms up to $p=4$ is 50. This might be a minimum criteria for determining the cutoff radius $(R_c)$ of the interatomic potentials in the $AB$-type compounds with guaranteed accuracy. Interestingly, Seko {\it et al.} have constructed an interatomic potential for the bcc K with $a=5.284$ \AA, and shown that the energy, force, and stress calculated by using the interatomic potential agree well with those obtained by the DFT calculations when the value of $R_c$ is set to be more than 9 \AA \ \cite{seko}. Such a $R_c$ is quite similar to $\sqrt{11}a/2=8.762$ \AA. 

Some compounds have a large discrepancy of the lowest phonon energy between the 6NN model and the DFPT. For example, the CdAg, MnHg, ZnCu, ZnAg, and ZnAu systems have $(\omega_{1},\omega_{\rm DFPT})=(-3.8,-0.9)$, $(-2.0,1.5)$, $(-5.7,-1.7)$, $(-3.5,3.0)$, and $(-3.2,2.4)$, respectively, in units of meV. This implies that more long-range interactions are required to achieve the convergence to the DFPT results. It should be noted that the stability of the $d$ electron compounds might not be described accurately within the PBE approximation. The value of $E_{\rm form}$ tends to be overestimated within the PBE when one studies the B2 compound that consists of the atoms having the completely filled $d$ orbitals such as Cu, Ag, Au, Zn, and Cd \cite{ruzsinszky2020}. More analysis using other functionals is beyond the scope of this work. The comparisons between the $p$NN and the DFPT for the 416 compounds are provided in the Supplemental Material \cite{SM}.




{\it Conclusions.} In conclusion, we studied an origin of the dynamical stability for 416 B2 compounds extracted from the MP \cite{MP}, and identified that the 266 out of 416 compounds are dynamically stable. The lowest phonon energy at the $M$ point determines whether the B2 compound is dynamically stable, and the interatomic interactions up tp the 4NN atoms are necessary for an accurate description of the B2 phase stability, which should determine the minimum cutoff radius for constructing the interatomic potentials for the $AB$-type compounds. The instability of the other compounds may be attributed to (i) the positive formation energy of the B2 structure (such as CuN) and/or (ii) the presence of different structures (such as CsF).


The present work assumes the B2 ordering on bcc structure. A competition between the B2 ordering and the finite temperature effects play an important role to understand the stability of binary \cite{turchi} and high-entropy alloys \cite{neugebauer}. Exploring a phonon stabilization under such situations is left for future work. In addition, the role of the spin-orbit coupling on the metastability \cite{schonecker} needs to be studied when heavy elements are included. We hope that this work stimulates other studies involving different crystal structures and gives a useful insight for developing the interatomic potentials of binary systems. 

\begin{acknowledgments}
This work was supported by JSPS KAKENHI (Grant No. JP21K04628). A part of numerical calculations has been done using the facilities of the Supercomputer Center, the Institute for Solid State Physics, the University of Tokyo.
\end{acknowledgments}




\begin{thebibliography}{99}

\bibitem{davey} W. P. Davey and F. G. Wick, The crystal structure of two rare halogen salts, Phys. Rev. {\bf 17}, 403 (1921). 



\bibitem{sluiter} M. H. F. Sluiter, Some observed bcc, fcc, and hcp superstructures, Phase Transitions {\bf 80}, 299 (2007).

\bibitem{hart2007} G. L. W. Hart, Where are nature's missing structures?, Nat. Mater. {\bf 6}, 941 (2007).



\bibitem{kolli} S. K. Kolli, A. R. Natarajan, J. C. Thomas, T. M. Pollock, and A. Van der Ven, Discovering hierarchies among intermetallic crystal structures, Phys. Rev. Mater. {\bf 4}, 113604 (2020).

\bibitem{MP} A. Jain, S. P. Ong, G. Hautier, W. Chen, W. D. Richards, S. Dacek, S. Cholia, D. Gunter, D. Skinner, G. Ceder, and K. A. Persson, The Materials Project: A materials genome approach to accelerating materials innovation, APL Mater. {\bf 1}, 011002 (2013).

\bibitem{sims} C. E. Sims, G. D. Barrera, N. L. Allan, and W. C. Mackrodt, Thermodynamics and mechanism of the $B1$-$B2$ phase transition in group-I halides and group-II oxides, Phys. Rev. B {\bf 57}, 11164 (1998).

\bibitem{florez} M. Fl\'orez, J. M. Recio, E. Francisco, M. A. Blanco, and A. Mart\'{\i}n Pend\'as, First-principles study of the rocksalt--cesium chloride relative phase stability in alkali halides, Phys. Rev. B {\bf 66}, 144112 (2002).



\bibitem{ono_koba} S. Ono and D. Kobayashi, Lattice stability of ordered Au-Cu alloys in the warm dense matter regime, Phys. Rev. B {\bf 103}, 094114 (2021).


\bibitem{khenata} R. Khenata and M. Sahnoun and H. Baltache and M. R\'{e}rat and D. Rached and M. Driz and B. Bouhafs, Structural, electronic, elastic and high-pressure properties of some alkaline-earth chalcogenides: An ab initio study, Physica B: Condens. Matter {\bf 371}, 12 (2006). 

\bibitem{wangB2} X. F. Wang, Travis E. Jones, W. Li, and Y. C. Zhou, Extreme Poisson's ratios and their electronic origin in B2 CsCl-type AB intermetallic compounds, Phys. Rev. B {\bf 85}, 134108 (2012).  

\bibitem{alsalmi} O. Alsalmi, M. Sanati, R. C. Albers, T. Lookman, and A. Saxena, First-principles study of phase stability of bcc $X\mathrm{Zn}$ ($X=\mathrm{Cu}$, Ag, and Au) alloys, Phys. Rev. Materials {\bf 2}, 113601 (2018).

\bibitem{grimvall} G. Grimvall, B. Magyari-K\"ope, V. Ozoli\ifmmode \mbox{\c{n}}\else \c{n}\fi{}\ifmmode \check{s}\else \v{s}\fi{}, and K. A. Persson, Lattice instabilities in metallic elements, Rev. Mod. Phys. {\bf 84}, 945 (2012).


\bibitem{frank} W. Frank, C. Els\"{a}sser, and M. F\"{a}hnle, Ab initio Force-Constant Method for Phonon Dispersions in Alkali Metals, Phys. Rev. Lett. {\bf 74}, 1791 (1995).

\bibitem{persson} K. Persson, M. Ekman, and V. Ozoli\ifmmode \mbox{\c{n}}\else \c{n}\fi{}\ifmmode \check{s}\else \v{s}\fi{}, Phonon instabilities in bcc Sc, Ti, La, and Hf, Phys. Rev. B {\bf 61}, 11221 (2000).

\bibitem{ono2019} S. Ono, Lattice dynamics for isochorically heated metals: A model study, J. Appl. Phys. {\bf 126}, 075113 (2019).


\bibitem{adam} J. B. Adams and S. M. Foiles, Development of an embedded-atom potential for a bcc metal: Vanadium, Phys. Rev. B {\bf 41}, 3316 (1990). 

\bibitem{seko} A. Seko, A. Takahashi, and I. Tanaka, First-principles interatomic potentials for ten elemental metals via compressed sensing, Phys. Rev. B {\bf 92}, 054113 (2015).

\bibitem{zuo} Y. Zuo, C. Chen, X. Li, Z. Deng, Y. Chen, J. Behler, G. Cs\'{a}nyi, A. V. Shapeev, A. P. Thompson, M. A. Wood, and S. P. Ong, Performance and cost assessment of machine learning interatomic potentials, J. Phys. Chem. {\bf 124}, 731 (2020). 


\bibitem{pymatgen} S. P. Ong, W. D. Richards, A. Jain, G. Hautier, M. Kocher, S. Cholia, D. Gunter, V. L. Chevrier, K. A. Persson, and G. Ceder, Python Materials Genomics (pymatgen): A robust, open-source python library for materials analysis, Comput. Mater. Sci. {\bf 68}, 314 (2013).




\bibitem{qe} P. Giannozzi, O. Andreussi, T. Brumme, O. Bunau, M. B. Nardelli, M. Calandra, R. Car, C. Cavazzoni, D. Ceresoli, and M. Cococcioni {\it et al}., Advanced capabilities for materials modeling with Quantum ESPRESSO, J. Phys.: Condens. Matter {\bf 29}, 465901 (2017).


\bibitem{pbe} J. P. Perdew, K. Burke, and M. Ernzerhof, Generalized Gradient Approximation Made Simple, Phys. Rev. Lett. {\bf 77}, 3865 (1996).


\bibitem{dalcorso} A. Dal Corso, Pseudopotentials periodic table: From H to Pu, Computational Material Science {\bf 95}, 337 (2014).

\bibitem{MK} H. J. Monkhorst and J. D. Pack, Special points for Brillouin-zone integrations, Phys. Rev. B {\bf 13}, 5188 (1976).

\bibitem{smearing} N. Marzari, D. Vanderbilt, A. De Vita, and M. C. Payne, Thermal Contraction and Disordering of the Al(110) Surface, Phys. Rev. Lett. {\bf 82}, 3296 (1999).
 
\bibitem{dfpt} S. Baroni, S. Gironcoli, A. Dal Corso, and P. Giannozzi, Phonons and related crystal properties from density-functional perturbation theory, Rev. Mod. Phys. {\bf 73}, 515 (2001).  
 


\bibitem{hart} G. L. W. Hart, S. Curtarolo, T. B. Massalski, and O. Levy, Comprehensive Search for New Phases and Compounds in Binary Alloy Systems Based on Platinum-Group Metals, Using a Computational First-Principles Approach, Phys. Rev. X {\bf 3}, 041035 (2013).

\bibitem{FeRh} J. M. Lommel and J. S. Kouvel, Effects of mechanical and thermal treatment on the structure and magnetic transitions in FeRh, J. Appl. Phys. {\bf 38}, 1263 (1967). 


\bibitem{SM} See Supplemental Material at (URL) for the derivation of Eqs.~(\ref{eq:omegaI}) and (\ref{eq:Cq}), the list of the 416 compounds, the phonon dispersions of the 416 compounds, and the $M$ point phonon energies within the $p$NN models and the DFPT calculations for the 416 compounds. 


\bibitem{VRu1} R. M. Waterstrat and R. C. Manuszewski, The vanadium-ruthenium constitution diagram, J. Less-Common Met. {\bf 48}, 151 (1976). 

\bibitem{VRu2} M. Marezio, P. D. Dernier, and C. W. Chu, Low-temperature X-Ray diffraction studies of near-equiatomic VRu alloys, Phys. Rev. B {\bf 4}, 2825 (1971). 


\bibitem{sun} W. Sun, S. T. Dacek, S. P. Ong, G. Hautier, A. Jain, W. D. Richards, A. C. Gamst, K. A. Persson, and G. Ceder, The thermodynamic scale of inorganic crystalline metastability, Sci. Adv. {\bf 2}, e1600225 (2016).

\bibitem{ono_meta} S. Ono, Metastability relationship between two- and three-dimensional crystal structures: a case study of the Cu-based compounds, Sci. Rep. {\bf 11}, 14588 (2021). 

\bibitem{wang} B. Wang and G. Frapper, Prediction of two-dimensional Cu$_2$C with polyacetylene-like motifs and Dirac nodal line, Phys. Rev. Mater. {\bf 5}, 034003 (2021).

\bibitem{ono_simple} S. Ono, Dynamical stability of two-dimensional metals in the periodic table, Phys. Rev. B {\bf 102}, 165424 (2020).

\bibitem{ono_satomi} S. Ono and H. Satomi, High-throughput computational search for two-dimensional binary compounds: Energetic stability versus synthesizability of three-dimensional counterparts, Phys. Rev. B {\bf 103}, L121403 (2021).


\bibitem{maradudin} A. A. Maradudin, E. W. Montroll, G. H. Weiss, and I. P. Ipatova, {\it Theory of Lattice Dynamics in the Harmonic Approximation} (Academic Press, New York, 1971).





\bibitem{ruzsinszky2020} N. K. Nepal, S. Adhikari, B. Neupane, and A. Ruzsinszky, Formation energy puzzle in intermetallic alloys: Random phase approximation fails to predict accurate formation energies, Phys. Rev. B {\bf 102}, 205121 (2020).


\bibitem{turchi} P. E. A. Turchi, M. Sluiter, F. J. Pinski, D. D. Johnson, D. M. Nicholson, G. M. Stocks, and J. B. Staunton, First-Principles Study of Phase Stability in Cu-Zn Substitutional Alloys, Phys. Rev. Lett. {\bf 67}, 1779 (1991).


\bibitem{neugebauer} F. K\"{o}rmann, T. Kostiuchenko, A. Shapeev, and J. Neugebauer, B2 ordering in body-centered-cubic AlNbTiV refractory high-entropy alloys, Phys. Rev. Materials {\bf 5}, 053803 (2021). 

\bibitem{schonecker} S. Sch\"{o}necker, X. Li, M. Richter, and L. Vitos, Lattice dynamics and metastability of fcc metals in the hcp structure and the crucial role of spin-orbit coupling in platinum, Phys. Rev. B {\bf 97}, 224305 (2018). 

\end{thebibliography}
\end{document}